\documentclass[12pt,preprint]{aastex}

\usepackage{lscape}
\usepackage[dvipdfm]{hyperref}

\shorttitle{Search for Point-Like Sources}
\shortauthors{The Telescope Array Collaboration}

\begin{document}



\title{A Northern Sky Survey for Point-Like Sources of EeV Neutral Particles with the Telescope Array Experiment}


\author{
R.U.~Abbasi$^{1}$,
M.~Abe$^{13}$,
T.Abu-Zayyad$^{1}$, 
M.~Allen$^{1}$, 
R.~Anderson$^{1}$, 
R.~Azuma$^{2}$, 
E.~Barcikowski$^{1}$, 
J.W.~Belz$^{1}$, 
D.R.~Bergman$^{1}$, 
S.A.~Blake$^{1}$
% ...
}
\affil{
$^{1}$ High Energy Astrophysics Institute and Department of Physics and Astronomy, University of Utah, Salt Lake City, Utah, USA \\
$^{2}$ Graduate School of Science and Engineering, Tokyo Institute of Technology, Meguro, Tokyo, Japan \\
% ...
$^{13}$ The Graduate School of Science and Engineering, Saitama University, Saitama, Saitama, Japan \\
% ...
}




\begin{abstract}
We report on the search for steady point-like sources of neutral
particles around 10$^{18}$~eV between 2008 May and
2013 May with the scintillator surface detector of the Telescope Array
experiment.  We found overall no significant point-like excess above
0.5~EeV in the northern sky.  Subsequently, we also searched for
coincidence with the {\it Fermi} bright Galactic sources.  No significant coincidence was
found within the statistical uncertainty.
Hence, we set an upper limit on the neutron
flux that corresponds to an averaged flux of 0.07 km$^{-2}$~yr$^{-1}$
for $E>1$~EeV in the northern sky at the 95\% confidence level.  This is the most stringent flux
upper limit in a northern sky survey assuming point-like
sources.  The upper limit at the 95\% confidence level on the neutron flux
from Cygnus~X-3 is also set to 0.2 km$^{-2}$~yr$^{-1}$ 
for $E>0.5$~EeV.  This is an order of magnitude lower than previous flux
measurements.
\end{abstract}


\keywords{acceleration of particles --- cosmic rays --- surveys}



\section{Introduction} \label{s-1}

\sloppy

The energy region around 10$^{18}$~eV (EeV) is thought to be a
transition from cosmic rays of galactic origin to those of extragalactic
origin.  Many cosmic-ray experiments have searched for point-like
sources as the origin of cosmic rays on the isotropic cosmic-ray sky in
this energy region.  Among them, the Fly's Eye experiment and the
Akeno 20-km$^{2}$ array independently reported a point-like excess
around Cygnus~X-3 above 0.5~EeV with a statistical significance at the
3~$\sigma$ level \citep{Cas89,Tes90}. In contrast, the Haverah
Park array found no significant excess around Cygnus~X-3 during a
period that overlaps most of the Fly's Eye observation
\citep{Law89}.  After these observations, however, there has been no
systematic search of the northern sky in this energy region to date.
The HiRes collaboration did search for point-like deviations from isotropy
in the northern sky for $E>10^{18.5}$~eV, however this is an order of
magnitude higher than the energy threshold of the previous Cygnus~X-3
observations \citep{Abb07}.  Recently, the Pierre Auger Observatory (PAO) surveyed
for point-like sources around 1~EeV with large statistics in the
southern sky.  They concluded that there was no significant
excess, although the stacked cosmic-ray events from the directions of 10
{\it Fermi} bright sources showed a potential excess of 2.35~$\sigma$ above
1~EeV \citep{dOrf11}.  The energy flux limits set by the PAO are
well below those observed from some Galactic TeV gamma-ray sources.
Therefore, they infer that this indicates that TeV gamma-ray emission
from those sources might be of electromagnetic origin, or their
proton spectra do not extend up to EeV energies \citep{Abr12}.

There are a few possibilities regarding the particle types and
distances of the point-like sources at EeV energies.  The mean free
path length of gamma rays with an energy of EeV is estimated to be approximately $330\times E^{0.9}$~kpc,
which strongly limits them to the neighborhood of our galaxy.
The mean decay length of
neutrons with an energy of EeV is calculated to be $9.2\times E$~kpc, which
corresponds to the Galactic center distance at 1~EeV.  Neutrons
with $E>2$~EeV enable us to look out over all of our Galaxy.  The
Larmor radius of protons with an energy of EeV is estimated to be approximately $0.3\times
E$~kpc at 3~$\mu$G within our Galaxy.  This curvature, however, makes it impossible
to find point-like sources.  Consequently, neutrons and gamma rays
from Galactic sources are the
most promising in the search for point-like sources.

The Telescope Array (TA) experiment has been observing
ultra-high-energy cosmic rays with $E\gtrsim10^{18}$~eV since 2008.  We
are probing the origins of ultra-high-energy cosmic rays using the
observational results from the TA, such as the cosmic-ray energy spectrum, mass
composition, and directional anisotropy. Our current results are
summarized as follows. The detailed energy spectrum above 10$^{18.2}$~eV was measured
\citep{Abu13a,Abu13b,Abu14a}, and it shows a steepening at
$5.7\times10^{19}$~eV, which is consistent with theoretical 
expectation from the Greisen-Zatsepin-Kuzmin (GZK) cutoff \citep{Gre66,Zat66}.
The preliminary result for the cosmic-ray composition
above 10$^{18.2}$~eV was consistent with the proton prediction within
the statistical and systematic uncertainties \citep{Tam13}.  
We also put stringent upper limits on the
absolute flux of ultra-high-energy photons at energies 
$E>10^{19}$~eV \citep{Abu13e}.  These limits strongly constrain
top-down models on the origin of cosmic rays.
TA has searched for UHECR anisotropies such as autocorrelations,
correlations with AGNs, and correlations with the LSSs of the
universe using the first 40 months of SD data \citep{Abu12a,Abu13d}.
Using the 5-year SD data, we updated results of the cosmic-ray anisotropy 
for $E>57$~EeV, which show deviations from isotropy at the significance
of 2--3~$\sigma$ \citep{Fuk13}. Finally, we observe on indication for 
large-scale anisotropy of cosmic rays with $E>57$~EeV in the northern hemisphere sky 
using the 5-year data set with additional statistics collected with the SD \citep{Abb14}.
The probability of this anisotropy appearing by chance in an isotropic
cosmic-ray sky is calculated to be 3.7$\times$10$^{-4}$ (3.4~$\sigma$). 

In this paper, we
report on the search for point-like sources of neutral particles, such
as neutrons or photons, at relatively low energies,
$E>5\times10^{17}$~eV with the high cosmic-ray statistics from the surface
detector of the TA experiment, which has the largest effective area in
the northern hemisphere.

\section{Experiment} \label{s-2}

The TA is the largest cosmic-ray detector in the northern hemisphere
and consists of a surface detector (SD) array \citep{Abu12a} and
three fluorescence detector (FD) stations \citep{Tok12}. The TA
has been in full operation in Millard Country, Utah, USA
($39\fdg30$N, $112\fdg91$W; about 1,400~m above sea level) since
2008. The TA SD array consists of 507 plastic scintillator
counters, each 3~m$^2$ in area, placed at grid points 1.2~km apart; it
covers an area of approximately 700~km$^{2}$. The TA SD array observes
cosmic rays with $E\gtrsim0.5$~EeV, regardless of the weather
conditions, using the extensive air shower (EAS) technique with a
duty cycle of 24 hours and a wide field of view (FoV). These
capabilities ensure a very stable and large geometrical exposure for
the northern sky survey, in comparison with the FD observations, for which
the duty cycle is limited to $\sim$10\%.

\section{SD Air Shower Analysis} \label{s-3}

The air shower reconstruction and data selection were optimized for the
low-energy air showers around $10^{18}$~eV on the basis of the 
reconstruction method developed in anisotropy and energy spectrum
studies \citep{Abu12a,Abu13a,Abu14b}.
To measure an accurate energy spectrum with the EAS
technique, the absolute acceptance of the EAS array as a function of the
energy must be carefully determined.
Therefore, air shower reconstruction usually
requires the elimination of reconstructed of events which lack
excellent energy resolution.
However, the absolute acceptance is not always required in this analysis
because we deduce the cosmic-ray backgrounds
from the data themselves by the equi-zenith angle method, as described in the following sections.
Hence, we substantially loosen the event cuts in the air shower reconstruction, at the cost of
good energy resolution.
The number of events remaining in the reconstruction used in the energy spectrum study \citep{Abu13a}
is relatively small ($\sim$14,800 events above 10$^{18.2}$~eV) owing to many hard parameter cuts,
which is called the ``standard cut'' in this paper,
mainly to improve the energy resolution for spectrum study.
To search for small- and large-scale anisotropy, air shower
statistics is more important than energy resolution if the
anisotropy changes gradually with the energy.
Table~\ref{tbl-1} shows the number of remaining events according to four simple
criteria that are defined as the loose cuts: (1) Each event must include at least four scintillator counters; (2) the zenith angle
of the event arrival direction must be less than 55$\degr$; (3) the
angular uncertainty estimated by the timing fit must be less than
10$\degr$; (4) the reconstructed energy must be greater than 0.5~EeV.
The number of triggered events is $\sim10^{6}$.
The trigger condition is the three-fold coincidence of adjacent SD elements with greater than three vertical equivalent muons within 8~$\mu$s
\citep{Abu12b}.
The number of air showers after the loose cuts is $\sim$10 times larger around 1~EeV
compared with that in the ``standard-cut'' data.
This is a remarkable advantage in the search for anisotropy in the EeV energy region,
even though the angular resolution and energy resolution are moderately degraded.
The angular resolution with the loose-cut data is estimated to be
$3\fdg0$ for $E>1$~EeV, whereas that of the standard-cut data is estimated to be $2\fdg2$.
The energy resolution with the loose-cut data is estimated to be $^{+50}_{-35}$\%,
whereas that of the standard-cut data is estimated to be $\sim^{+35}_{-25}$\%.

The optimization of the air shower reconstruction for low-energy air showers
was studied by a Monte Carlo (MC) simulation
based on CORSIKA version 6.960 \citep{Cor98}, with hadronic interaction models
QGSJET-II-03, FLUKA2008.3c, and EGS4 \citep{Nel85} for air shower event generation
and the GEANT4 for the response of each scintillator counter \citep{Sto12}.
Primary cosmic rays are generated on the basis of
the energy spectrum measured by the HiRes experiment at energies of
10$^{17.2}$~eV to 10$^{20.4}$~eV \citep{Abb08}. Because of the uncertainty in the composition of primary cosmic rays
in this energy region,
we use pure proton and pure iron in this MC simulation. Further, the core locations of
simulated air shower events are uniformly distributed over a
circle 25~km in radius and centered at the central laser facility
\citep{Udo07}, which is located in the center of the TA at a distance of 20.85~km from each of the FD stations. 
These simulated events were analyzed in the same way as the experimental data
to deduce the energy and arrival direction of cosmic
rays, including the detailed detector responses and calibrations such as
the dead time of detectors and time variations in the detector gains.

In this analysis, we re-optimized the geometric reconstruction of the arrival direction
using the modified Linsley time-delay 
function \citep{Tes86}:
\begin{equation}\label{Eq1}
T_{\rm d} = a \left(1+\frac{r}{30} \right)^{1.5} \rho^{0.5},
\end{equation}
where $T_{\rm d}$ is the time delay of air shower particles from the
shower plane (ns), $r$ is the perpendicular distance from the shower axis
(m), $\rho$ is the pulse height per unit area (VEM/m$^{2}$,
where VEM is the vertical equivalent muon, which is the average pulse height
produced by vertically penetrating muons in the detector), and $a$ is
the Linsley curvature parameter \citep{Lin62}. 
The curvature parameter ``$a$'' was a free parameter in the previous analysis \citep{Abu13a}.
However, the number of misreconstructions increase for the low-energy
air showers which were detected by the small number of detectors.
Therefore, the $a$ set to be fixed parameter to reduce the misreconstructions,
and optimized as $a(\theta)= 2.2~{\rm cos}(1.1~\theta)$ by
the MC simulation dependence on the zenith angle $\theta$.

The energy was estimated from a lateral distribution fit with the same
form as that used in the standard-cut analysis \citep{Abu13a}.
First, we calculate $S(800)$, the density of air shower particles
at a lateral distance of 800~m from the core, by the lateral distribution fit.
Then, $S(800)$ was converted to the energy using a look-up table
for $S(800)$ and the zenith angle determined from the MC simulation using the loose-cut events.
The energies reconstructed by the SD were renormalized by 1/1.27
to match the SD energy scale to that of the FD, which was determined calorimetrically \citep{Abu13a}.

\begin{figure}
\epsscale{0.50}
\plotone{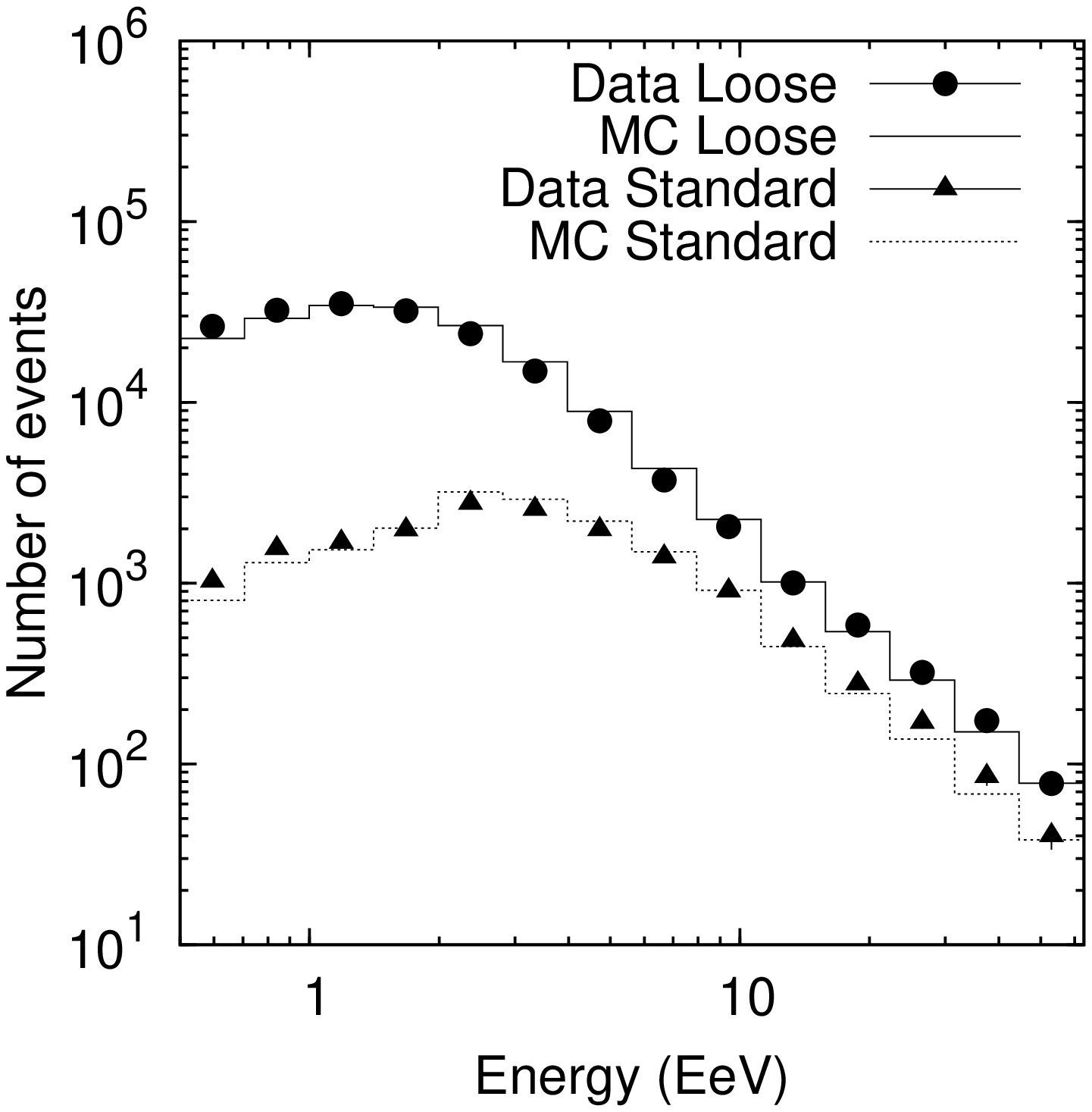}
\caption{ Reconstructed energy distributions.  Closed circles and triangles
  show energy distributions of experimental data after loose and
  standard cuts, respectively.  Solid and dashed histograms show
  energy distributions in MC simulation after loose and standard
  cuts, respectively. The areas of MC simulation are normalized to those of the data.}\label{fig1}
\end{figure}

Figure~\ref{fig1} shows the reconstructed energy distribution and compares data to MC.
The MC simulation is consistent with the data distribution.
In this figure, one can see that the reconstruction efficiency
with the loose cuts around 1~EeV is increased by 10 times
compared with that with the standard cuts.  This is a remarkable advantage
in the search for anisotropy in the EeV energy region.
For the point-like
source search, we divided the loose-cut data set into four energy
regions: $0.5<E ({\rm EeV})\le1.0$ (58,895 events),
$1.0<E ({\rm EeV})\le2.0$ (67,277 events), $E ({\rm EeV})>2.0$ (54,472
events), and $E ({\rm EeV})>1.0$ (121,749 events).
The first energy threshold, 0.5~EeV, corresponds to the energy of
the Cygnus X-3 fluxes measured by the Akeno array and the Fly's Eye.
The data set with the highest energy threshold extends the range to visible neutron sources anywhere in our galaxy.

\section{Hybrid Data Analysis} \label{s-4}

\begin{figure}
\epsscale{0.50}
\plotone{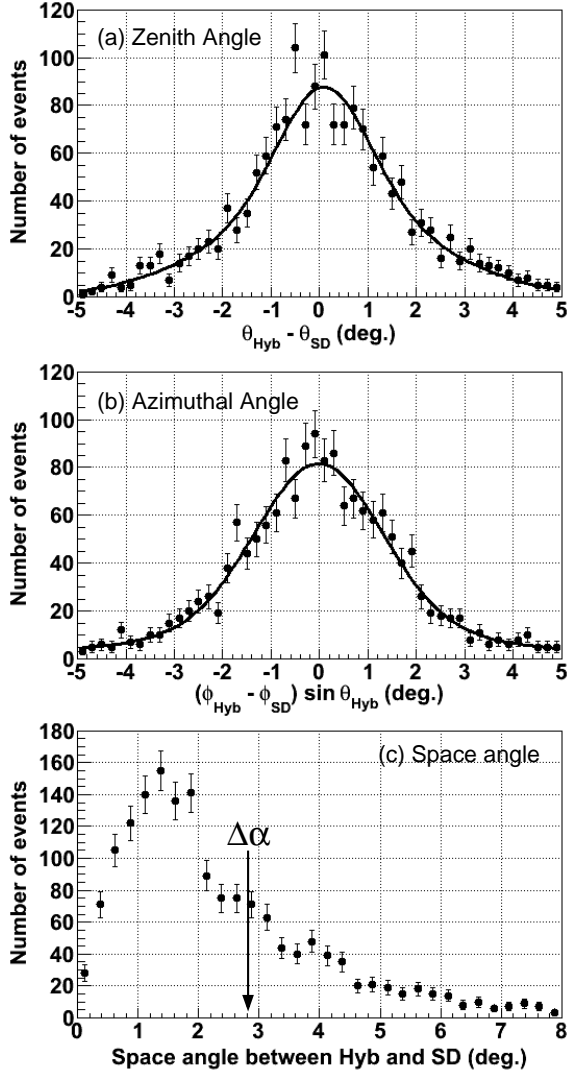}
\caption{Opening angle distributions of directions measured by
  hybrid method and SD.  Solid curves represent best fits
  by double-Gaussian function with the common mean
  value. (a) Opening angle distributions of zenith angles measured by
  hybrid ($\theta_{\rm Hyb}$) method and SD ($\theta_{\rm SD}$). Estimated mean
  opening angle is $m=+0\fdg091\pm0\fdg046$. (b) Opening angle
  of azimuthal angles measured by
  hybrid ($\phi_{\rm Hyb}$) method and SD ($\phi_{\rm SD}$). Estimated mean opening
  angle is $m=-0\fdg022\pm0\fdg046$. (c) Space angle distribution
  of direction measured by the hybrid method and the SD. The vertical arrow indicates 
  a space angle containing 68\% of the events ($\Delta\alpha = 2\fdg8\pm0\fdg1$).
} \label{fig2}
\end{figure}

The performance of the SD was thoroughly verified by a FD-SD
hybrid data analysis \citep{Abu14a} independent of the MC simulation.
The arrival directions of the hybrid events were
determined by the fluorescence track
measured by the FD and the air shower arrival position at the ground measured by the SD.
This hybrid reconstruction is almost independent
of the SD reconstruction, and its angular resolution, $\sigma_{\rm Hyb}=1\fdg0\pm0\fdg1$, 
is better than that of the SD reconstruction, $\sigma_{\rm SD}\sim3\degr$ around 1~EeV.
Therefore, the hybrid data are a good reference for estimating the systematic errors in the arrival direction.
Figure~\ref{fig2}(a) shows the opening angle distributions of
the zenith angles measured using the hybrid method ($\theta_{\rm Hyb}$) and the SD ($\theta_{\rm SD}$).
Figure~\ref{fig2}(b) shows the opening angle distributions of
the azimuthal angles measured using the hybrid method ($\phi_{\rm Hyb}$) and the SD ($\phi_{\rm SD}$).
The solid curves are fitted to the data points by a double-Gaussian function,
\begin{equation}\label{Eq2}
G(\delta) = g_{1}(\delta;a_{1}, m, \sigma_{1}) + g_{2}(\delta;a_{2}, m, \sigma_{2}),
\end{equation}
where $g_{i}(\delta) = a_{i} e^{-(\delta-m)^{2}/2\sigma_{i}^{2}}$, $i$ indicates the $i$-th Gaussian function,
$a_{i}$ is the $i$-th height, $m$ is the common mean value in the two Gaussians,
$\sigma_{i}$ is the $i$-th standard deviation, and $\delta$ is the opening angle.
The mean opening angles of the zenith angle is calculated to be $m = +0\fdg091\pm0\fdg046$
($\sigma_{1}=2\fdg16\pm0\fdg12$, $\sigma_{2}=0\fdg92\pm0\fdg10$) using the Chi-squared minimization technique,
while that of the azimuthal angle is calculated to be $m = -0\fdg022\pm0\fdg046$ 
($\sigma_{1}=3\fdg63\pm1\fdg64$, $\sigma_{2}=1\fdg30\pm0\fdg11$).
From these results, the systematic pointing error of the reconstructed SD shower is estimated to be approximately $0\fdg1$.
This is obviously negligible compared with our angular resolution.
Figure~\ref{fig2} (c) shows the distribution of the space angle between the
directions measured by SD and the hybrid method above 0.5~EeV. A space angle
containing 68\% of the events $\Delta\alpha$ is estimated to be $2\fdg8\pm0\fdg1$.

\begin{figure}
\epsscale{0.50}
\plotone{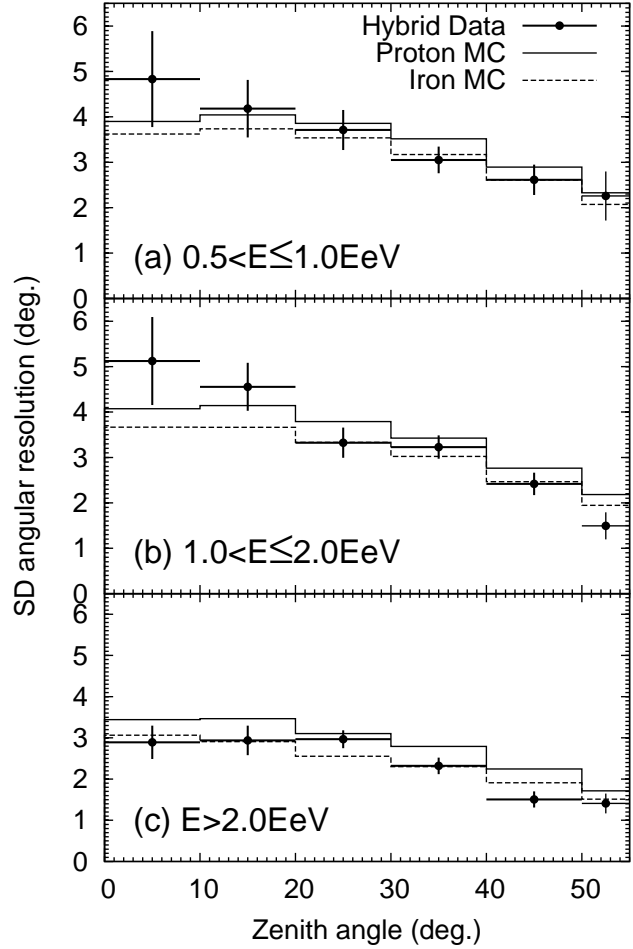}
\caption{ Angular resolutions of TA SD as a function of zenith
  angle.  Solid histograms show results of MC simulation
  assuming pure protons.  Dashed histograms show results of MC
  simulation assuming pure iron.  Closed circles indicate the
  estimated angular resolution from opening angles of
  reconstructed directions between SD and hybrid data.
  Zenith angle of hybrid data is limited to $55\degr$.
  MC simulation and hybrid data are divided into three
  energy regions: (a) $0.5 < E {\rm (EeV)} \le 1.0$ (upper panel), (b)
  $1.0 < E {\rm (EeV)} \le 2.0$ (middle panel), (c) $ E {\rm (EeV)} >
  2.0$ (lower panel).  }\label{fig3}
\end{figure}

We studied the angular resolution of the cosmic-ray arrival directions
containing 68\% of the reconstructed events, dependence on the zenith angle.
In Figure~\ref{fig3}, the solid and dashed histograms show the angular
resolutions of the MC simulations for proton and iron, respectively.
The closed circles show the estimated angular resolution $\sigma_{\rm
  SD}$ from the following quadratic sum relation.
\begin{equation}\label{Eq3}
\sigma_{\rm SD}=\sqrt{\Delta\alpha^{2} - \sigma_{\rm Hyb}^{2}},
\end{equation}
where, $\Delta\alpha$ is a space angle, containing 68\% of the events,
between the directions measured by the SD and the hybrid method, which corresponds to 
Figure~\ref{fig2} (c), $\sigma_{\rm Hyb}=1\fdg0\pm0\fdg1$ is assumed to be the angular resolution of
the hybrid method independent of the zenith angle and energy \citep{Abu14a}.  The
$\sigma_{\rm SD}$ values estimated from the SD-FD hybrid data are in reasonable agreement
with the MC simulation results. They agree to better than 3~$\sigma$ for all energy bins.
Above 2~EeV, the angular resolution estimated
from the hybrid data is slightly better than that from the proton MC, whereas it is consistent with that from the iron MC.
Because the composition of primary cosmic rays in this energy region
is still under debate owing to the systematic uncertainty,
the average difference in angular resolution between the data and the proton MC
is defined as a systematic error of the angular resolution, which corresponds to $\sim$15\%, assuming the worst case.
Thus, these estimations from the hybrid data are good checks for the
reconstruction of the TA SD independent of the MC simulation.
In Figure~\ref{fig3}, the angular resolution is clearly improved at larger
zenith angles, and the iron-induced air shower shows slightly
better resolution than the proton-induced air shower. This is because the
footprint of the air shower at large zenith angles is larger than that at small
zenith angles, and the muon component is much greater at large zenith angles
than at small zenith angles. The time distribution of muons in
an air shower is narrower than that of the electromagnetic components.
Therefore, air showers with a high ratio of muons to secondary particles,
such as large-zenith-angle or iron-induced air showers, might
enable us to better determine the geometry of the air shower front.

\begin{figure}
\epsscale{0.50}
\plotone{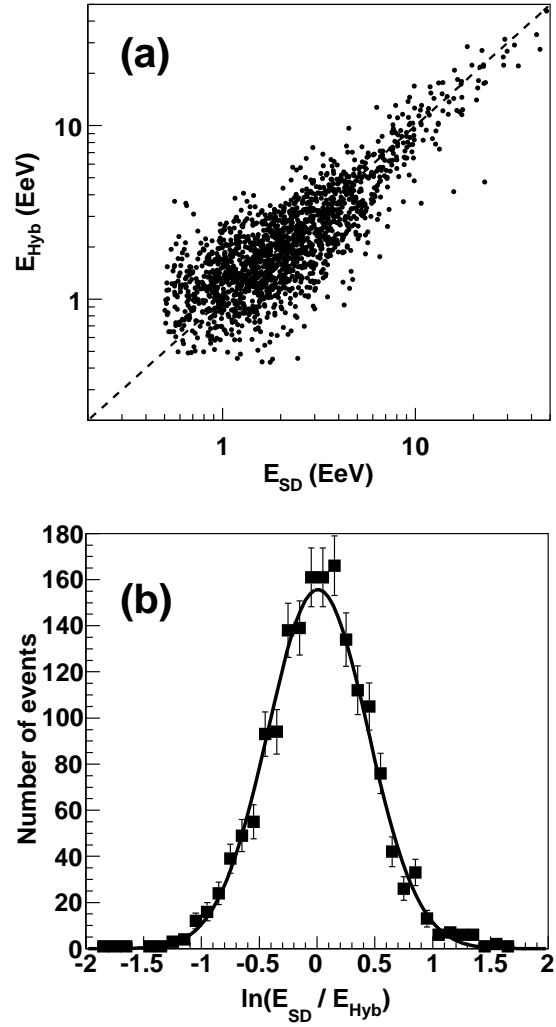}
\caption{ Comparison of reconstructed energies from SD and hybrid method.
(a) Scatter plot of reconstructed energy from SD and hybrid method.
(b) Natural logarithm of ratio of reconstructed energy from SD and hybrid method.
Energy resolution with loose-cut data is estimated to be $^{+50}_{-35}$\%.
}\label{fig4}
\end{figure}

Finally, we estimate the energy resolution of the TA SD with the loose-cut data set using the hybrid data.
Figure~\ref{fig4}(a) shows a scatter plot of the reconstructed energy
from the SD and the hybrid method. Figure~\ref{fig4}(b) shows the distribution
of the natural logarithm of the ratio of the reconstructed energy from the SD and the hybrid method.
The energy resolution of the hybrid analysis is 7\%, which is sufficiently better than that of the SD \citep{Abu14a}.
From Figure~\ref{fig4}, the energy resolution of the SD with the loose-cut data is estimated to be $\sim$$^{+50}_{-35}$\%
for $E>1$~EeV, whereas that with the standard-cut data is estimated to be $\sim$$^{+35}_{-25}$\%.
This resolution is good enough to find a point-like source
if its flux changes gradually with the energy.

\section{Background Calculation} \label{s-5}

Various background estimation methods have been developed to
analyze the cosmic-ray anisotropy.  A simple method is to compare the
distribution of air shower directions generated by the MC simulation
directly with the data.
In this analysis, the typical number of background events for a target source is up to $\sim$200.
To determine the significance of the excess within an accuracy of 0.1~$\sigma$,
the background should be estimated as 0.7\% ($=\sqrt{200}\times0.1/200$).
However, the MC simulation usually does not reproduce the data with this accuracy
due to the simulation model dependence and
meteorological effects, which are difficult to incorporate into the MC
simulation.  In the alternative method, the background can be estimated by
the data themselves without the MC simulation. To extract an excess of air
shower events coming from the direction of a target source, we adopt
the equi-zenith angle method developed by the Tibet AS$\gamma$
experiment \citep{Ame03} to find gamma-ray excesses from huge
cosmic-ray background events in the TeV energy region.  The signals
are searched for by counting the number of events coming from a target
source in an on-source cell with a finite size.  The background is
estimated by the number of events averaged over six off-source cells
with the same angular radius as the on-source cell at the same zenith
angle, recorded at the same time as the on-source cell
events. Note that the equi-zenith angle method fails when the
source object stays at a zenith angle of less than 10$\degr$, because the off-source cells
overlap with other cells. Therefore, the air shower events with zenith angles
larger than 10$\degr$ were used in this analysis.

\begin{figure}
\epsscale{0.50}
\plotone{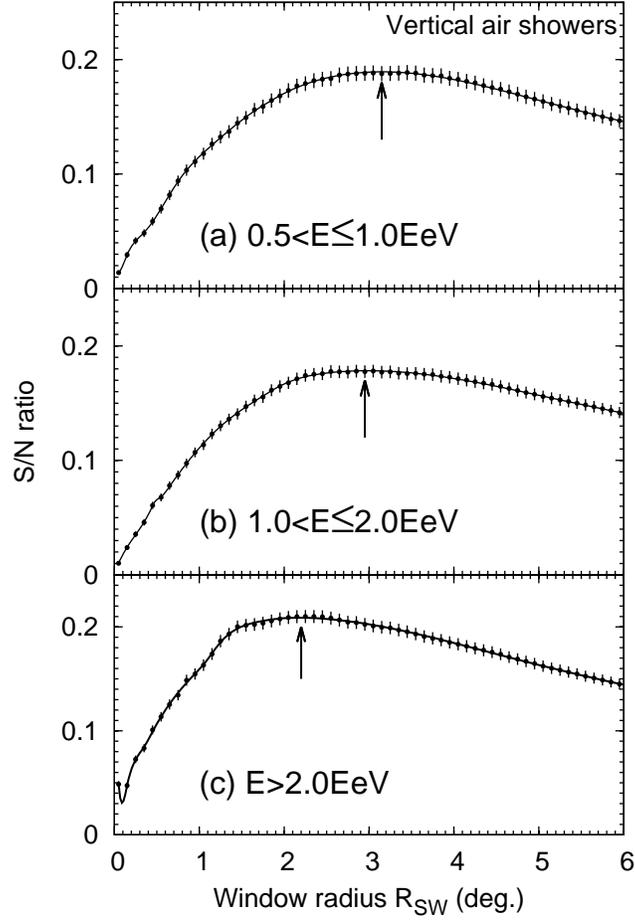}
\caption{ S/N curves for vertical air showers ($\theta < 5\degr$)
  as a function of search window size $R_{\rm sw}$ from MC
  simulation in three energy regions: (a) $0.5 < E {\rm (EeV)} \le 1.0$, (b)
  $1.0 < E {\rm (EeV)} \le 2.0$, (c) $ E {\rm (EeV)} > 2.0$.
  The solid curves are a spline fitting to the MC data as shown by the points.
  Noise ($=\sqrt{B}$, square root of number of background events) is
  proportional to the area of the search window $\pi R_{\rm sw}^{2}$.  Peak
  (solid arrow) indicates optimal search window radius to maximize
  S/N ratio.  }\label{fig5}
\end{figure}

\begin{figure}
\epsscale{0.50}
\plotone{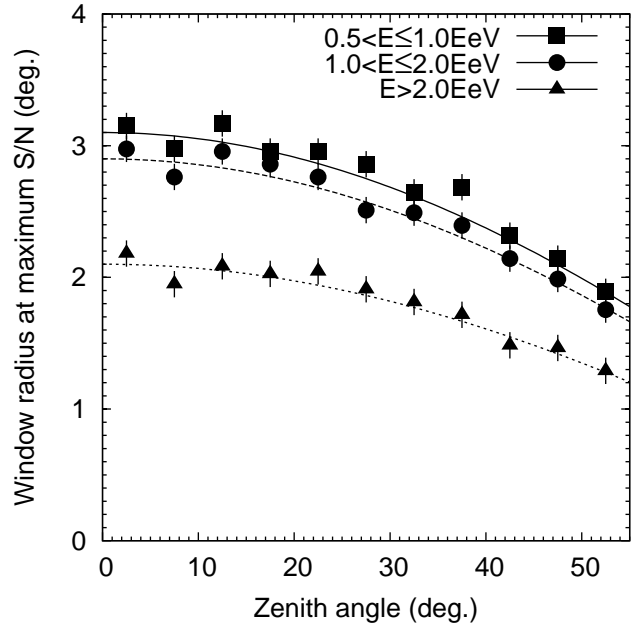}
\caption{ Optimal search window radius as a function of zenith angle
  $\theta$ from MC simulation.  The curves show the best fit by the
  empirical formula $R_{\rm sw} = R_{0}~{\rm cos}\theta$.  Symbols
  and line types represent theree different energy regions:
  squares with solid curve, $0.5 < E {\rm (EeV)} \le 1.0$; circles with
  dotted curve, $1.0 < E {\rm (EeV)} \le 2.0$; triangles with dashed
  curve, $ E {\rm (EeV)} > 2.0$.  }\label{fig6}
\end{figure}

The search window size of the on- and off-source cells should be optimized
by the MC simulation to maximize the signal-to-noise (S/N) ratio where the signal is the number of detected
excess events, and the noise is the square root of the number of background events ($\sqrt{B}$), which
depends on the angular resolution. In this MC study, we generated air
showers induced by protons, which have the same air shower development
as those induced by neutrons.  Figure~\ref{fig5} shows S/N ratio as a
function of the search window radius $R_{\rm sw}$ in the MC simulation.
The vertical axis is the S/N ratio, defined as the number of signals
divided by $R_{\rm sw}$ ($=\sqrt{R_{\rm sw}^{2}}$), assuming that the
the number of background events ($B$) is proportional to the area of the search window $\pi
R_{\rm sw}^{2}$.  A peak (arrow position) indicates the optimal search
window radius to maximize the S/N ratio.  Figure~\ref{fig6} shows the
optimal search window radius $R_{\rm sw}$ at the maximum S/N ratio as a function of the zenith
angle $\theta$ and the results of fitting by an empirical formula, $R_{\rm
  sw}(\theta) = R_{0} {\rm cos} \theta$, where $R_{0}$
is the fitting parameter denoting the window radius for a vertical
air shower. The calculated $R_{0}$ values are $3\fdg1$, $2\fdg9$, and $2\fdg1$ for
three energy regions: $0.5 < E {\rm (EeV)} \le 1.0$,
  $1.0 < E {\rm (EeV)} \le 2.0$, and $ E {\rm (EeV)} > 2.0$, respectively.
In this analysis, we use these fitting curves in Figure~\ref{fig6} as
the optimal search window radius.
If the signals show a normal Gaussian distribution,
the optimal window size $R_{\rm sw}(\theta)$ should be close to the angular resolution $\sigma_{\rm SD}$.
$R_{\rm sw}(\theta)$ is, however,  $0.6\sim0.7$ times smaller than $\sigma_{\rm SD}$
because the signal spread shows a large-tail distribution.

The off-source cells are located
in the azimuthal direction at the same zenith angle as the on-source direction.
Four off-source cells are symmetrically aligned on
each side of the on-source cell, at $6\fdg4$ steps from the on-source position measured in terms of the real angle,
and pick up events recorded at the same time to the on-source cell.
This method, the so-called
equi-zenith angle method, can reliably estimate the background
events under the same conditions as those for the on-source events.  Here, it is
worth noting that the two off-source cells adjacent to the on-source
cell are excluded to avoid possible signal tail leakage into the
off-source events.  Therefore, the total number of off-source cells is six. The
TA SD has the anisotropy of 6\% at the maximum in the azimuthal
direction owing to the azimuthal dependence of the trigger efficiency.
This anisotropy, which is well understood, appears along the grid 
of detector arrangement for the air showers with small number of hit detectors. 
To correct this anisotropy, we analyzed 19 dummy sources,
which follow the same diurnal rotation (at the same declination and a spacing of 18$\degr$ in right ascension, except for the location of the object itself) in the same way as for
the target source using the equi-zenith angle method.  The background
distribution of the mean values of the observed air shower events for the 19
dummy sources reproduces the background shape of the object at the same declination
very well \citep{Ame03}.
The number of events in the $i$-th off-source cell of the target
source $n_{\rm off}^{i}$ is corrected by the number of events at the
on- and off-source cells averaged over the 19 dummy sources using the following
equation:
\begin{equation}\label{Eq4}
N_{\rm off}^{i} = n_{\rm off}^{i} \left( \frac{\langle D_{\rm on}\rangle}{\langle D_{\rm off}^{i}\rangle} \right),
\end{equation}
where $N_{\rm off}^{i}$ is the corrected number of events at the $i$-th
off-source cell, $\langle D_{\rm off}^{i}\rangle$ is the number of
events averaged over the 19 dummy sources at the $i$-th off-source cell, and
$\langle D_{\rm on}\rangle$ is the number of events averaged over the 19
dummy sources at the on-source cell.  This correction enables us to
remove the anisotropy of off-source events completely if the azimuthal
anisotropy is stable. This is because the 19 dummy sources are observed on a
different part of the sky every day, as they all orbit with the same
diurnal rotation. The correction factors $\langle D_{\rm on}\rangle / \langle D_{\rm off}^{i}\rangle$
are $1.0\pm0.03$ which depends on the declination.

Finally, we calculate the statistical significance of cosmic-ray
signals from the target sources against cosmic-ray background events
using the following formula \citep{Li83}:
\begin{equation}\label{Eq5}
S_{\rm LM} = \sqrt{2}\left[N_{\rm on}~{\rm ln}\left( \frac{(1+\eta)N_{\rm on} }{\eta (N_{\rm on} + N_{\rm off})}\right) + N_{\rm off}~{\rm ln}\left(\frac{(1+\eta) N_{\rm off} }{N_{\rm on} + N_{\rm off}} \right)\right]^{1/2},
\end{equation}
where $N_{\rm on}$, $N_{\rm off}$, and $\eta$ are the number of
events in the on-source cell, the number of corrected background events summed
over six off-source cells ($N_{\rm off}=\sum_{i=1}^{6} N_{\rm off}^{i}$), and the ratio of the on-source solid angle area
to the off-source solid angle area ($\eta$ = 1/6 in this work),
respectively.

\section{Results and Discussion} \label{s-6}

\begin{landscape}
\begin{figure}
\epsscale{1.0}
\plotone{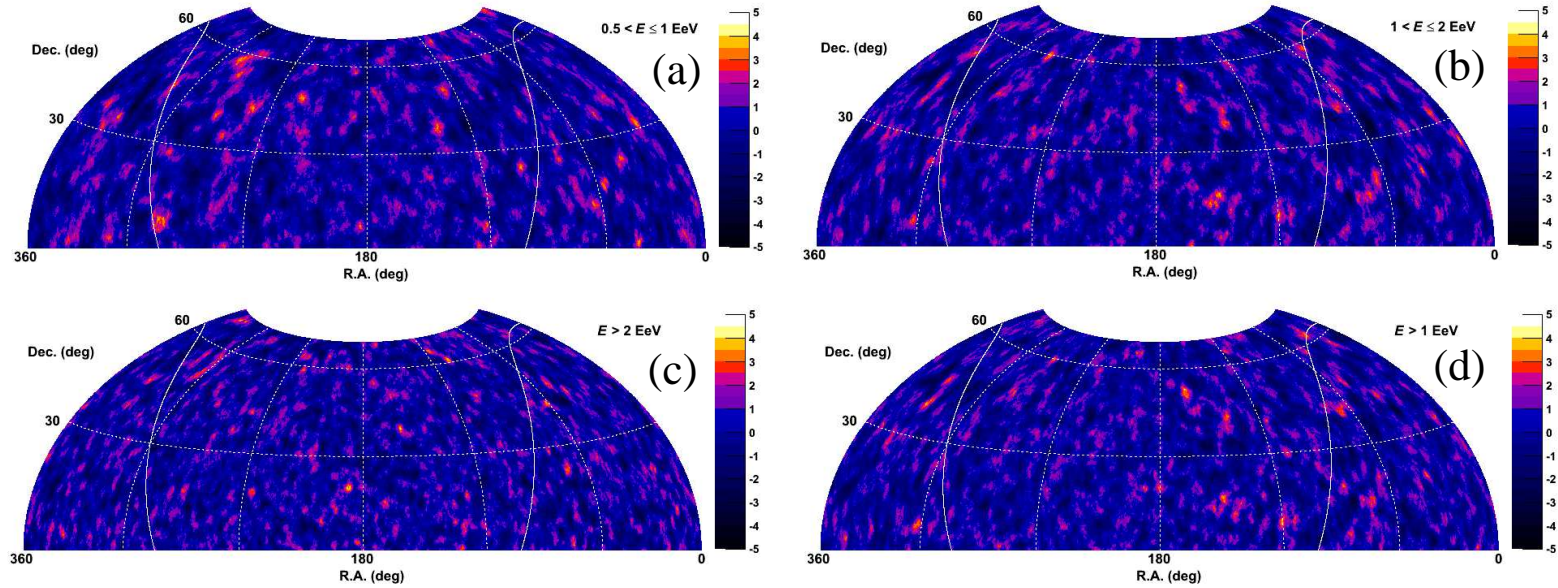}
\caption{ Significance maps of the northern sky between
  Dec. = 0$\degr$ and Dec. = $70\degr$ surveyed by TA SD in four energy regions: (a)
  $0.5 < E {\rm (EeV)} \le 1.0$, (b) $1.0 < E {\rm (EeV)} \le 2.0$,
  (c) $ E {\rm (EeV)} > 2.0$, (d) $ E {\rm (EeV)} > 1.0$.  Color
  contours show significance level.  Solid curves indicate
  the Galactic plane.  }\label{fig7}
\end{figure}
\end{landscape}

\begin{figure}
\epsscale{0.80}
\plotone{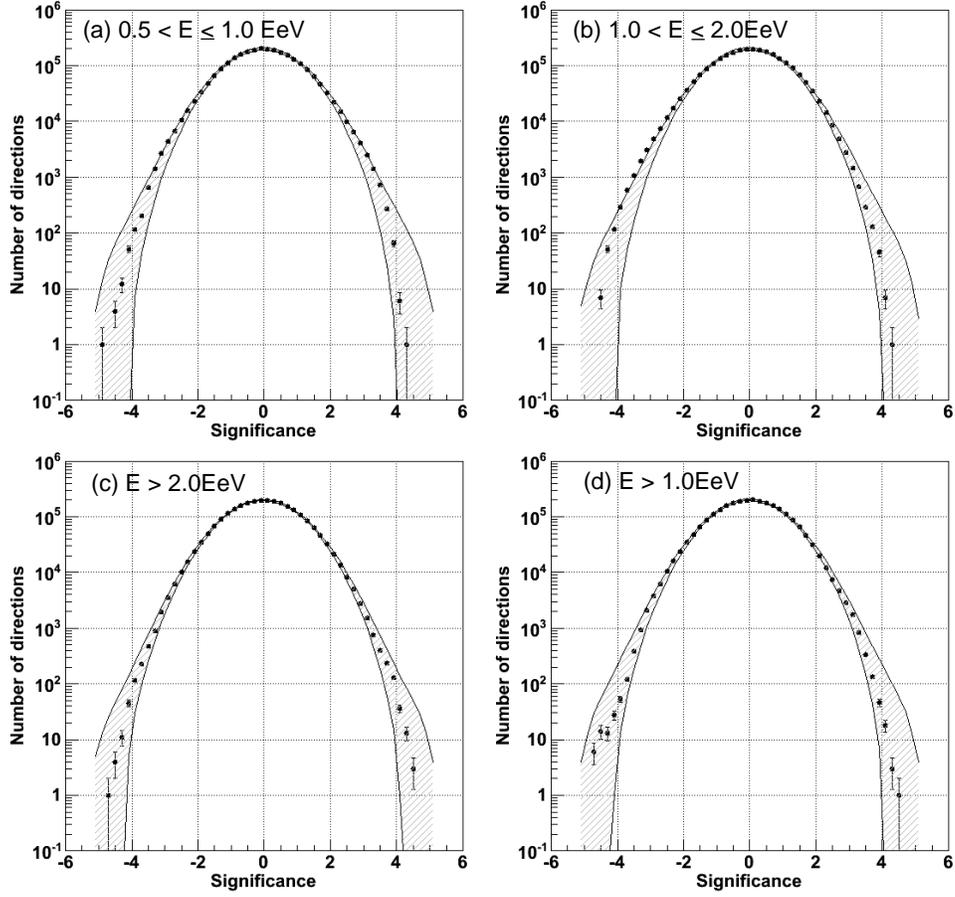}
\caption{ Histograms showing significance distributions in all directions
  within FoV of TA SD in four energy regions: (a) $0.5 < E {\rm (EeV)}
  \le 1.0$, (b) $1.0 < E {\rm (EeV)} \le 2.0$, (c) $ E {\rm (EeV)} >
  2.0$, (d) $ E {\rm (EeV)} > 1.0$.  Shaded area indicates 95\%
  containment region of 10$^5$ MC samples of isotropic sky.
}\label{fig8}
\end{figure}

We analyze 180,644 air showers collected by the TA SD from 2008 May
11 to 2013 May 4.
Figure~\ref{fig7} shows the northern significance sky map drawn by
the equi-zenith angle method using cosmic rays observed by the TA SD
in the four energy regions.  In this analysis, to ensure that we did not miss
any possible unknown sources, the surveyed sky was
oversampled. The centers of the tested target sources are set on $0\fdg1
\times 0\fdg1$ grids, from 0$\degr$ to 360$\degr$ in right ascension and from
0$\degr$ to 70$\degr$ in declination.  At each grid point, a search window
with the optimal radius $R_{\rm sw}(\theta)$, as shown in
Figure~\ref{fig6}, was opened.  The observed declination band is limited by
the statistics and the analysis method.  The number of events at
Dec.$ < 0\degr$ is small.  At Dec. $> 70\degr$, near the
northern pole, the dummy source cells overlap other cells, so
the statistical independence of each cell fails.  The
closed circles in Figure~\ref{fig8} show the significance distributions
from all directions in the four energy regions.  The shaded area is
the 95\% containment region of 10$^{5}$ MC samples in
the isotropic sky.  The good agreement between the data points and shaded area
indicates that there is overall no significant excess beyond
the statistical fluctuation in the northern sky.

\begin{figure}
\epsscale{0.80}
\plotone{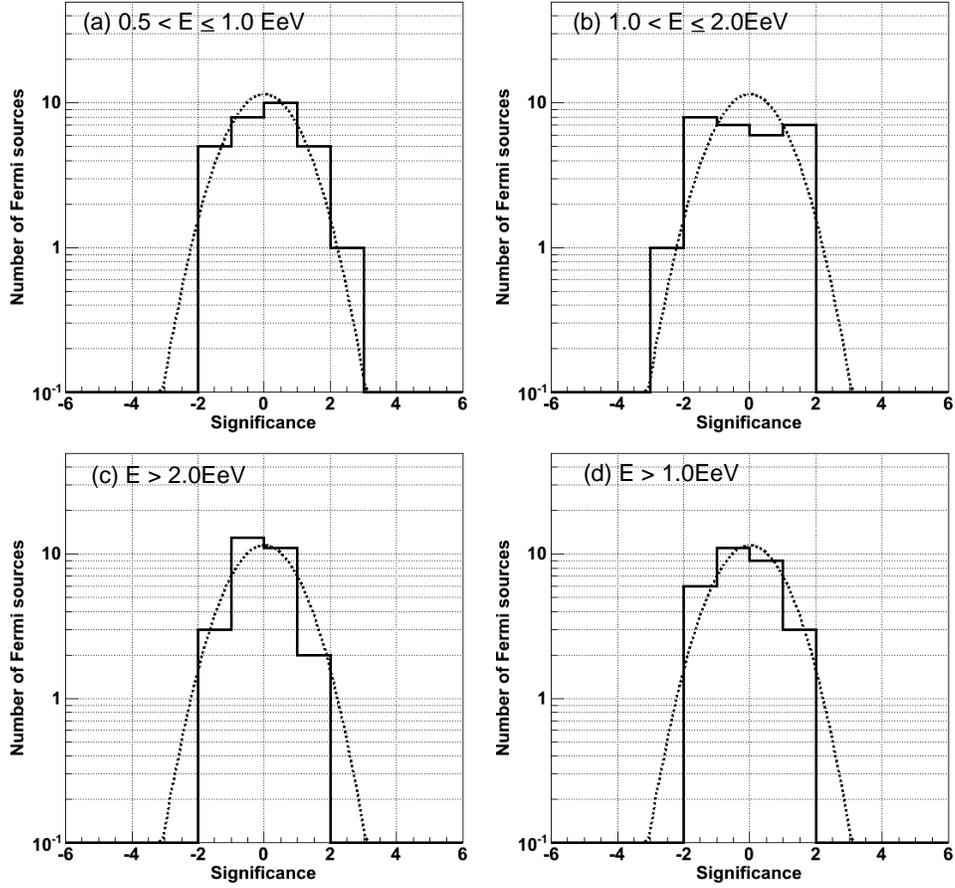}
\caption{ Histograms showing significance distributions of 29 {\it Fermi}
  bright Galactic sources within FoV of TA SD in four energy regions: (a)
  $0.5 < E {\rm (EeV)} \le 1.0$, (b) $1.0 < E {\rm (EeV)} \le 2.0$,
  (c) $ E {\rm (EeV)} > 2.0$, (d) $ E {\rm (EeV)} > 1.0$.  Dotted
  curves are expected normal Gaussian distributions.
}\label{fig9}
\end{figure}

Subsequently, we also searched for coincidence with the {\it Fermi}
bright Galactic sources.  The target sources in the {\it Fermi} bright
source list \citep{Abd09a} were chosen as confirmed or potential
Galactic sources in the same way as for the TeV
observations of the Milagro and Tibet AS$\gamma$ \citep{Abd09b,Ame10}.
Out of the 205 most significant sources in the {\it Fermi} bright
source list, 84 are not identified as extragalactic sources.  Among
these 84, we selected 29 sources in the declination band between
0$\degr$ and 70$\degr$, corresponding to the sensitive
FoV of the TA SD. The results of the search for neutral particles
from the 29 {\it  Fermi} bright Galactic sources are summarized in Table~\ref{tbl-2},
where 15 of the selected sources are classified as pulsars (PSR), 5
are supernova remnants (SNR), 1 is a high-mass X-ray binary (HXB), and 8
remain unidentified but are potential Galactic sources; they are
mostly concentrated in the Galactic plane ($|b|<\sim20\degr$)
\citep{Abd09a}.  Many {\it Fermi} bright Galactic sources are confirmed
sources at TeV energies, as shown in Table~\ref{tbl-2}.
Figure~\ref{fig9} shows the significance distributions of the 29
{\it Fermi} source directions searched by the TA SD. The distributions are obviously consistent with
the normal Gaussian distribution, indicating that there are no statistically significant signals from these sources.

\begin{figure}
\epsscale{0.5}
\plotone{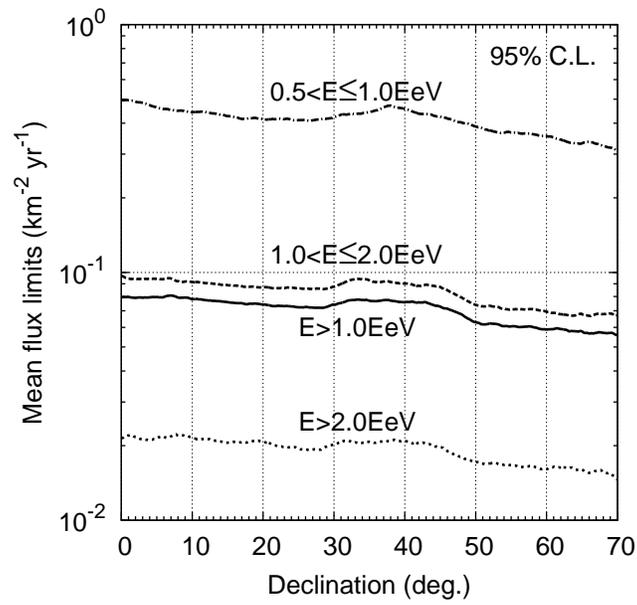}
\caption{ Mean flux upper limits (km$^{-2}$ yr$^{-1}$) at 95\% C.L.
  according to the declination of the target sources observed by TA SD
  at each energy.  Dot-dashed curve: $0.5 < E {\rm (EeV)} \le 1.0$; dashed
  curve: $1.0 < E {\rm (EeV)} \le 2.0$; dotted curve: $ E {\rm (EeV)}
  > 2.0$; solid curve: $ E {\rm (EeV)} > 1.0$.  }\label{fig10}
\end{figure}

We calculated the flux upper limits on the neutron intensity ($F_{\rm
  ul}$) of the northern sky using the following equation:
\begin{equation}\label{Eq6}
F_{\rm ul} = F_{\rm cr} \frac{N_{\rm ul}}{N_{\rm bg}} \frac{\omega_{\rm sw}}{\epsilon_{\rm sw}},
\end{equation}
where $F_{\rm cr}$ is the integral cosmic-ray flux;
$N_{\rm ul}$ is the upper limit
on the observed excess $(N_{\rm on} -N_{\rm bg})$ according to a
statistical prescription assuming an unphysical region, such as a region of negative
excess \citep{Hel83}; $N_{\rm bg} (= \eta N_{\rm off}$) is the average number
of background events; $\omega_{\rm sw}$ is the averaged solid
angle of the search window for a target source depending on the declination; and $\epsilon_{\rm sw}$ is the signal efficiency
with the angular cut by $\omega_{\rm sw}$ deduced from the MC simulation of proton ($\sim$ neutron)
assuming a point source.
The $F_{\rm cr}$ values at $0.5$~EeV, $1$~EeV, and $2$~EeV are assumed
to be fluxes measured by HiRes \citep{Abb08}\footnote{\url{http://www.physics.rutgers.edu/~dbergman/HiRes-Monocular-Spectra-200702.html}} 
because the TA spectrum below 10$^{18.2}$~eV has not been published yet.  The HiRes
spectrum is consistent with that of the TA within 5\% at
10$^{18.2}$~eV. The value of $\epsilon_{\rm sw}$
is estimated to be $0.50\pm0.01$ for energies between 0.5~EeV and 1.0~EeV,
and $0.45\pm0.01$ for $E>1$~EeV, independent of the declination of the target source.
The typical fractions of the upper excess ($N_{\rm ul} / N_{\rm bg}$) in
each energy bin are 29\% for $0.5 < E {\rm (EeV)} \le 1.0$, 29\% for
$1.0 < E {\rm (EeV)} \le 2.0$, 46\% for $ E {\rm (EeV)} > 2.0$, and
25\% for $ E {\rm (EeV)} > 1.0$.
First, we calculated the flux upper limit of the entire northern sky
point by point on $0\fdg1 \times 0\fdg1$ grids using Eq.~\ref{Eq6}.
Then, the mean of the flux limits at the same declination was
defined as the representative value at each declination.
Figure~\ref{fig10} shows the representative mean flux upper limits
(km$^{-2}$ yr$^{-1}$) at the 95\% C.L. in the four energy
regions according to the declination of the target sources.  The average
flux upper limit in the northern sky is estimated to be 0.07 km$^{-2}$
yr$^{-1}$ for $E>1$~EeV.  This is the most stringent flux upper limit in
a northern sky survey assuming point-like sources.  The flux
upper limit for each {\it Fermi} bright Galactic source is also listed
in Table~\ref{tbl-2}.  

The fluxes of Cygnus~X-3 reported by the Fly's Eye and the Akeno
20-km$^{2}$ array are $(2.0\pm0.6)\times10^{-17}$ cm$^{-2}$ s$^{-1}$
and $(1.8\pm0.7)\times10^{-17}$ cm$^{-2}$ s$^{-1}$ for $E>0.5$~EeV,
respectively \citep{Cas89,Tes90}.  Our observational results for
Cygnus~X-3 are summarized in Table~\ref{tbl-3}.  The upper limit
at the 95\% C.L. on the neutron flux of Cygnus~X-3 observed by the TA SD is
estimated to be 0.2~km$^{-2}$~yr$^{-1}$ (=$5.6\times10^{-19}$ cm$^{-2}$ s$^{-1}$) 
for $E>0.5$~EeV, as shown in Table~\ref{tbl-3}.
This is an order of magnitude smaller than the fluxes
measured by the Fly's Eye and the Akeno 20-km$^{2}$ array.  One
possible explanation of their signals around Cygnus~X-3 could be
transient emission during their observation periods.
We divided the data set between 2008 May 11 and 2013 May 4
into 18 periods (1 period $\sim$100~days) and searched for
transient signals from Cygnus~X-3. We found no significant excess
in these 18 periods.

\section{Summary} \label{s-7}
We search for steady point-like sources of neutral particles in the
EeV energy range observed by the TA SD, which has the largest effective area in the
northern sky.  
The data selection was loosen and tuned the reconstruction of
the arrival direction in this analysis. As a result, the number
of air showers with $E>0.5$~EeV, which corresponds to 180,644 events, was
$\sim$10 times larger than the original ``standard cut'' analysis around EeV energies.
To search for point-like
sources, the equi-zenith angle method was applied to these cosmic-ray
air showers taken by the TA SD between 2008 May and
2013 May.  We found no significant excess
for $E>0.5$~EeV in the northern sky.  Subsequently, we also searched for
coincidence with the {\it Fermi} bright Galactic sources.  No
significant coincidence was found within the statistical error.  Hence, we
set upper limits at the 95\% C.L. on the neutron flux, which is an
averaged flux of 0.07 km$^{-2}$ yr$^{-1}$ for $E>1$~EeV in the northern
sky.  This is the most stringent flux upper limit in a northern sky
survey assuming point-like sources.  The upper limit at the 95\%
C.L. on the neutron flux of Cygnus~X-3 is estimated to be 0.2
km$^{-2}$ yr$^{-1}$ for $E>0.5$~EeV.  This is an order of magnitude
lower than the previous flux measurements.

\acknowledgments 
The Telescope Array experiment is supported by the Japan
Society for the Promotion of Science through Grants-in-Aids for Scientific Research on Specially 
Promoted Research (21000002) ``Extreme Phenomena in the Universe Explored
by Highest Energy Cosmic Rays'' and for Scientific Research
 (19104006), and the Inter-University Research Program of
the Institute for Cosmic Ray Research; by the U.S. National
Science Foundation awards PHY-0307098, PHY-0601915, PHY-0649681,
PHY-0703893, PHY-0758342, PHY-0848320, PHY-1069280, PHY-1069286, PHY-1404495 and PHY-1404502; by the National Research Foundation
of Korea (2007-0093860, R32-10130, 2012R1A1A2008381, 2013004883); by
the Russian Academy of Sciences, RFBR grants 11-02-01528a and 13-02-01311a (INR), IISN project No. 4.4509.10 and Belgian
Science Policy under IUAP VII/37 (ULB). The foundations
of Dr. Ezekiel R. and Edna Wattis Dumke, Willard
L. Eccles and the George S. and Dolores Dore Eccles all
helped with generous donations. The State of Utah supported
the project through its Economic Development Board, and the
University of Utah through the Office of the Vice President
for Research. The experimental site became available through
the cooperation of the Utah School and Institutional Trust
Lands Administration (SITLA), U.S. Bureau of Land Management,
and the U.S. Air Force. We also wish to thank the people
and the officials of Millard County, Utah for their steadfast
and warm support. We gratefully acknowledge the contributions
from the technical staffs of our home institutions. An
allocation of computer time from the Center for High Performance
Computing at the University of Utah is gratefully acknowledged.

\clearpage

\clearpage

\clearpage

\clearpage

\clearpage

\clearpage

\clearpage

\clearpage

\clearpage

\clearpage

\clearpage

\begin{deluxetable}{lr}
\tabletypesize{\scriptsize} 
\tablecaption{ 
Loose-cut parameters in this analysis and the number of remaining events
\label{tbl-1}}
\tablewidth{0pt}
\tablehead{
\colhead{Cut parameters}& \colhead{\# of events} 
}
\startdata
\# of triggered events &  1,133,213\\
\tableline
\# of SDs $\ge 4$&   296,208\\
Pointing direction error $<10\degr$ & 290,603\\
Zenith angle $\theta < 55\degr$& 255,332\\
Energy $>$ 0.5~EeV & 180,644 
\enddata
\end{deluxetable}

\clearpage

\begin{deluxetable}{cccccccc}
\tabletypesize{\scriptsize} 
\tablecaption{Results of search for EeV neutral particles from the 29 {\it Fermi} Galactic sources \label{tbl-2}}
\tablewidth{0pt}
\tablehead{
\colhead{{\it Fermi} LAT} & &R.A. &Dec. & \colhead{$S_{\rm LM}$ ($>$1EeV)} & \colhead{$F_{\rm ul}$($>$1EeV)\tablenotemark{*}} &  &\colhead{Source}\\
\colhead{Source(0FGL)} & \colhead{Class} &  \colhead{(deg.)}  &  \colhead{(deg.)}  & \colhead{($\sigma$)}  & \colhead{(km$^{-2}$ yr$^{-1}$)} &  TeV $\gamma$&\colhead{Associations} 
}
\startdata
J0030.3+0450 & PSR & 7.6 & 4.8 & $-$0.15 & $<$0.07 & &\\    
J0240.3+6113 & HXB & 40.0 & 61.2 & $-$0.35 & $<$0.05 & Yes & \\  
J0357.5+3205 & PSR & 59.3 & 32.0 & $-$1.19 & $<$0.04 & &\\    
J0534.6+2201 & PSR & 83.6 & 22.0 & $-$0.36 & $<$0.06 & Yes &Crab \\  
J0617.4+2234 & SNR & 94.3 & 22.5 & +1.20 & $<$0.11 & Yes & IC 443 \\
J0631.8+1034 & PSR & 97.9 & 10.5 & $-$0.29 & $<$0.07 & & \\   
J0633.5+0634 & PSR & 98.3 & 6.5 & $-$0.29 & $<$0.06 & & \\   
J0634.0+1745 & PSR & 98.5 & 17.7 & +0.21 & $<$0.09 & & Geminga \\  
J0643.2+0858 & - & 100.8 & 8.9 & $-$0.99 & $<$0.05 & & \\   
J1830.3+0617 & - & 277.5 & 6.2 & +0.93 & $<$0.12 & &\\    
J1836.2+5924 & PSR & 279.0 & 59.4 & $-$0.76 & $<$0.04 & & \\   
J1855.9+0126 & SNR & 283.9 & 1.4 & +0.90 & $<$0.11 & &W44 \\   
J1900.0+0356 & - & 285.0 & 3.9 & +0.73 & $<$0.11 & \\    
J1907.5+0602 & PSR & 286.8 & 6.0 & +0.84 & $<$0.10 & Yes & \\  
J1911.0+0905 & SNR & 287.7 & 9.0 & $-$0.85 & $<$0.05 & Yes &G43.3$-$0.17 \\   
J1923.0+1411 & SNR & 290.7 & 14.1 & $-$1.31 & $<$0.04 & Yes & W51 \\ 
J1953.2+3249 & PSR & 298.3 & 32.8 & $-$1.54 & $<$0.04 & & \\   
J1954.4+2838 & SNR & 298.6 & 28.6 & +0.47 & $<$0.09 & & G65.1+0.6 \\  
J1958.1+2848 & PSR & 299.5 & 28.8 & +0.70 & $<$0.09 & & \\   
J2001.0+4352 & - & 300.2 & 43.8 & $-$0.76 & $<$0.05 & & \\   
J2020.8+3649 & PSR & 305.2 & 36.8 & $-$0.93 & $<$0.05 & Yes & \\  
J2021.5+4026 & PSR & 305.3 & 40.4 & $-$0.24 & $<$0.07 & & \\   
J2027.5+3334 & - & 306.8 & 33.5 & +0.77 & $<$0.11 & & \\   
J2032.2+4122 & PSR & 308.0 & 41.3 & $-$1.25 & $<$0.04 & Yes &\\   
J2055.5+2540 & - & 313.8 & 25.6 & +1.04 & $<$0.10 & & \\   
J2110.8+4608 & - & 317.7 & 46.1 & $-$1.63 & $<$0.03 & & \\   
J2214.8+3002 & PSR & 333.7 & 30.0 & +0.55 & $<$0.09 & &\\    
J2229.0+6114 & PSR & 337.2 & 61.2 & +1.13 & $<$0.09 & Yes &\\    
J2302.9+4443 & - & 345.7 & 44.7 & $-$1.50 & $<$0.03 & & 
\enddata
\tablenotetext{*}{Upper limits on the neutron flux at 95\% C.L.}
\end{deluxetable}

\begin{deluxetable}{lcccccccc}
\tabletypesize{\scriptsize} 
\tablecaption{ 
Summary of Cygnus~X-3 observations by the TA SD
\label{tbl-3}}
\tablewidth{0pt}
\tablehead{
\colhead{Energy(EeV)} &  \multicolumn{2}{c}{$E>0.5$} & & \multicolumn{2}{c}{$E>1.0$} & & \multicolumn{2}{c}{$E>2.0$}\\
    \cline{2-3}  \cline{5-6}  \cline{8-9}
    & \colhead{$S_{\rm LM}$} & \colhead{$F_{\rm ul}$\tablenotemark{*}} & & \colhead{$S_{\rm LM}$} & \colhead{$F_{\rm ul}$\tablenotemark{*}} & & \colhead{$S_{\rm LM}$} & \colhead{$F_{\rm ul}$\tablenotemark{*}} \\
Object    & \colhead{($\sigma$)} & \colhead{(km$^{-2}$ yr$^{-1}$)} & & \colhead{($\sigma$)} & \colhead{(km$^{-2}$ yr$^{-1}$)}&  & \colhead{($\sigma$)} & \colhead{(km$^{-2}$ yr$^{-1}$)} 
}
\startdata
Cygnus~X-3 & $-$1.04 & $<$0.2 & & $-$1.55 & $<$0.03 & & $-$0.24 & $<$0.02
\enddata
\tablenotetext{*}{Upper limits on the neutron flux at 95\% C.L.}
\end{deluxetable}


\begin{thebibliography}{}

\bibitem[Abbasi et al. 2007]{Abb07}
Abbasi, R. U., Abu-Zayyad, T., Amann, J. F., et al. 2007, Astropart. Phys., 27, 512

\bibitem[Abbasi et al. 2008]{Abb08}
Abbasi, R. U., Abu-Zayyad, T., Allen, M., et al. 2008, \prl, 100, 101101

\bibitem[Abbasi et al. 2014]{Abb14}
Abbasi, R. U., Abe, M., Abu-Zayyad, T., et al. 2014, \apjl, 790, L21

\bibitem[Abdo et al. 2009a]{Abd09a}
Abdo, A. A., Ackermann, M., Ajello, M., et al. 2009a, \apjs, 183, 46

\bibitem[Abdo et al. 2009b]{Abd09b}
Abdo, A. A., Allen, B. T., Aune, T., et al. 2009b, \apjl, 700, L127

\bibitem[Abreu et al. 2012]{Abr12}
Abreu, P., Aglietta, M., Ahlers, M., et al. 2012, \apj, 760, 148

\bibitem[Abu-Zayyad et al. 2012a]{Abu12a}
Abu-Zayyad, T., Aida, R., Allen, M., et al. 2012a, \apj, 757, 26

\bibitem[Abu-Zayyad et al. 2012b]{Abu12b}
Abu-Zayyad, T., Aida, R., Allen, M., et al, 2012b, NIM-A, 689, 87

\bibitem[Abu-Zayyad et al. 2013a]{Abu13a}
Abu-Zayyad, T., Aida, R., Allen, M., et al. 2013a, \apjl, 768, L1

\bibitem[Abu-Zayyad et al. 2013b]{Abu13b}
Abu-Zayyad, T., Aida, R., Allen, M., et al. 2013b, Astropart. Phys., 48, 16

\bibitem[Abu-Zayyad et al. 2014a]{Abu14a}
Abu-Zayyad, T., Aida, R., Allen, M., et al. 2014a, Astropart. Phys., in press (arXiv:1305.7273)

\bibitem[Abu-Zayyad et al. 2014b]{Abu14b}
Abu-Zayyad, T., Aida, R., Allen, M., et al. 2014b, Astropart. Phys., in press (arXiv:1403.0644)

\bibitem[Abu-Zayyad et al. 2013d]{Abu13d}
Abu-Zayyad, T., Aida, R., Allen, M., et al. 2013d, \apj, 777, 88

\bibitem[Abu-Zayyad et al. 2013e]{Abu13e}
Abu-Zayyad, T., Aida, R., Allen, M., et al. 2013e, \prd, 88, 112005

\bibitem[Amenomori et al. 2003]{Ame03}
Amenomori, M., Ayabe, S., Cui, S. W., et al. 2003, \apj, 598, 242

\bibitem[Amenomori et al. 2010]{Ame10}
Amenomori, M., Bi, X. J., Chen, D., et al. 2010, \apjl, 809, L6

\bibitem[Cassiday et al. 1989]{Cas89}
Cassiday, G. L., Cooper, R., Dawson, B. R., et al. 1989, \prl, 62, 383

\bibitem[d'Orfeuil et al. 2011]{dOrf11}
d'Orfeuil, B. R., \& The Pierre Auger Collaboration 2011, Proceedings of 32nd ICRC (Beijing), 2, 91 (arXiv:1107.4805)

\bibitem[Fukushima et al. 2013]{Fuk13}
Fukushima, M., Ivanov, D., Kawata, K., et al. 2013, Proceedings of 33rd ICRC (Rio de Janeiro), CR-EX (Id:1033), in press

\bibitem[Greisen 1966]{Gre66}
Greisen, K. 1966, \prl, 16, 748

\bibitem[Heck et al. 1998]{Cor98}
Heck, D., Knapp, J., Capdevielle, J. N., Shatz, G., \& Thouw, T. 1998, 
CORSIKA: A Monte Carlo Code to Simulate Extensive Air Showers (FZKA 6019)(Karlsruhe: Forschungszentrum Karlsruhe)

\bibitem[Helene 1983]{Hel83}
Helene, O. 1983, Nucl. Instrum. Methods Phys. Res., 212, 319

\bibitem[Lawrence et al. 1989]{Law89}
Lawrence, M. A., Prosser, D. C. \& Watson, A. A. 1989, \prl, 63, 1121

\bibitem[Li \& Ma 1983]{Li83}
Li, T.-P., \& Ma, Y.-Q. 1983, \apj, 272, 317

\bibitem[Linsley \& Scarsi 1962]{Lin62}
Linsley, J., \& Scarsi, L. 1962, Phys. Rev., 128, 2384

\bibitem[Nelson et al. 1985]{Nel85}
Nelson, W. R., Hirayama, H., \& Rogers, D. W. O. 1985, Report No. SLAC-0265

\bibitem[Stokes et al. 2012]{Sto12}
Stokes, B. T., Cady, R., Ivanov, D., Matthews, J. N. \& Thomson, G. B.  2012, Astropart. Phys., 35, 759

\bibitem[Tameda 2013]{Tam13}
Tameda, Y. \& The Telescope Array Collaboration 2013, Proceedings of 33rd ICRC (Rio de Janeiro), CR-EX (Id:512), in press

\bibitem[Teshima et al. 1986]{Tes86}
Teshima, M., Matsubma, Y., Hara, T., et al. 1986, J. Phys. G: Nucl. Phys., 12, 1097

\bibitem[Teshima et al. 1990]{Tes90}
Teshima, M., Matsubara, Y., Nagano, M., et al. 1990, \prl, 64, 1628

\bibitem[Tokuno et al. 2012]{Tok12}
Tokuno, H., Tameda, Y., Takeda, M., et al. 2012, NIM-A, 676, 54

\bibitem[Udo et al. 2007]{Udo07}
Udo, S., Allen, M., Cady, R., et al. 2007, 30th ICRC (Merida), Proceedings of 30th ICRC (Merida), Ed. R. Caballero, et al. (Universidad Nacional Autonoma de Mexico, Mexico City, Mexico), 5, 1021

\bibitem[Zatsepin \& Kuz'min 1966]{Zat66}
Zatsepin, G. T., \& Kuz'min, V. A. 1966, J. Exp. Theor. Phys. Lett., 4, 78

\end{thebibliography}
\end{document}